\documentclass[a4paper,twocolumn,10pt,unpublished]{quantumarticle}
\pdfoutput=1
\usepackage[utf8]{inputenc}
\usepackage[english]{babel}
\usepackage[T1]{fontenc}
\usepackage{amsmath,amsthm,amssymb}
\usepackage{hyperref}
\usepackage{adjustbox}
\usepackage{multirow}
\usepackage{float}
\usepackage{graphicx}
\usepackage[numbers, sort&compress]{natbib}
\usepackage[all]{hypcap}
\usepackage{comment}
\usepackage{tabularx}
\usepackage{color}
\usepackage{braket}
\usepackage{ragged2e}

\newcommand{\be}{\begin{equation}}
\newcommand{\ee}{\end{equation}}
\newcommand{\ba}{\begin{eqnarray}}
\newcommand{\ea}{\end{eqnarray}}

\newcommand{\addtuc}{School of Electrical and Computer Engineering, Technical University of Crete, Chania, Greece 73100}

\newcommand{\addcqt}{Centre for Quantum Technologies, National University of Singapore, 3 Science Drive 2, Singapore 117543}

\newcommand{\addihpc}{Institute of High Performance Computing, 1 Fusionopolis Way, \#16-16 Connexis, Singapore 138632}

\newcommand{\addangel}{AngelQ Quantum Computing, 531A Upper Cross Street \#04-95 Hong Lim Complex, Singapore 051531}

\newcommand{\addibm}{IBM Quantum, IBM Research Zurich, Säumerstrasse 4, 8803 Rüschlikon, Switzerland}

\begin{document}
\title{Computing Electronic Correlation Energies using Linear Depth Quantum Circuits}
\author{Chong Hian Chee}
\email{ch.chee@u.nus.edu}
\affiliation{\addcqt}

\author{Adrian M. Mak}
\affiliation{\addihpc}

\author{Daniel Leykam}
\affiliation{\addcqt}

\author{Panagiotis Kl. Barkoutsos}
\affiliation{\addibm}

\author{Dimitris G. Angelakis}
\email{dimitris.angelakis@nus.edu.sg}
\affiliation{\addcqt}
\affiliation{\addtuc}
\affiliation{\addangel}

\begin{abstract}
Efficient computation of molecular energies is an exciting application of quantum computing for quantum chemistry, but current noisy intermediate-scale quantum (NISQ) devices can only execute shallow circuits, limiting existing variational quantum algorithms, which require deep entangling quantum circuit ansatzes to capture correlations, to small molecules. Here we demonstrate a variational NISQ-friendly algorithm that generates a set of mean-field Hartree-Fock (HF) ansatzes using multiple shallow circuits with depth linear in the number of qubits to estimate electronic correlation energies via perturbation theory up to the second order. We tested the algorithm on several small molecules, both with classical simulations including noise models and on cloud quantum processors, showing that it not only reproduces the equilibrium molecular energies but it also captures the perturbative electronic correlation effects at longer bond distances. As fidelities of quantum processors continue to improve our algorithm will enable the study of larger molecules compared to other approaches requiring higher-order polynomial circuit depth.
\end{abstract}

\maketitle

\section{Introduction}

Quantum chemical methods are routinely used to compute molecular energies and properties, which in turn are used to interpret experimental results and predict chemical phenomena~\cite{helgakerRecentAdvancesWave2012}. Quantum computation for chemistry has received much attention recently, with an increasing number of problems in chemistry, physics, and materials science that have been proposed to be solvable by a quantum computer~\cite{bauer2020}. These quantum computers can efficiently represent electronic wavefunctions that scale at most linearly in system size, whereas their classical counterparts would require an exponential scaling of memory for an equivalent representation~\cite{steudtner2018}. In the wider computing context, quantum computing approaches and algorithms with polynomial or even exponential speedups are emerging as alternatives to classical computation methods. 

For example, Quantum Phase Estimation (QPE)~\cite{abrams1997, abrams1999} is an early quantum algorithm that provides an exponential speedup in predicting molecular energies given a good guess for corresponding electronic eigenstates~\cite{aspuru-guzikSimulatedQuantumComputation2005a}. However, the required quantum circuit depth grows polynomially with molecular system size, which is deemed too deep for current noisy intermediate-scale quantum (NISQ) devices~\cite{preskill2018}, even for modest sized molecules~\cite{elfvingHowWillQuantum2020}.

Major efforts in the development of NISQ-friendly algorithms for quantum chemistry have largely relied on the hybrid quantum-classical framework~\cite{mccleanTheoryVariationalHybrid2016}, which flexibly splits the computation between quantum and ordinary classical computers. Within this framework, hybrid algorithms will generate a quantum circuit ansatz and measure various properties required to estimate the molecular energies on a NISQ device, but will also optimize the parameters within the quantum ansatz on a classical computer to improve the accuracy of the estimates. There are many other proposals and approaches under this framework that aim to adapt to the limitations of current NISQ devices including Quantum Subspace Expansion~\cite{mccleanHybridQuantumclassicalHierarchy2017a} and Fermionic Monte Carlo~\cite{hugginsUnbiasingFermionicQuantum2022}. 

Variational quantum algorithms (VQA) are one popular approach~\cite{delgadoVariationalQuantumAlgorithm2021, cerezo2021} which has seen many experimental NISQ demonstrations in estimating molecular energies of small molecules~\cite{peruzzo2014, aruteHartreeFockSuperconductingQubit2020, kandala2017, omalley2016}, often the ground state energy, using various types of quantum circuit ansatz and classical optimizers. VQA predicts the molecular energies via minimization of a function which is variational within the domain of the parameters defined.

Despite many experimental demonstrations of VQA for predicting molecular energies, there are several issues which makes it challenging to have practical applications in large-scale molecular systems. First, existing VQA approaches still need deep entangling quantum circuit ansatzes, for which the circuit depth scales at least polynomially in the number of qubits $N$, so as to accurately represent the electronic wavefunction by incorporating electronic repulsive interactions and correlation effects~\cite{anandQuantumComputingView2022, evangelistaExactParameterizationFermionic2019}. Second, the number of classical parameters needed to sufficiently parameterize the Hilbert space of a molecular state, that is necessary to accurately estimate the energies, can be potentially huge, which makes ansatz optimization difficult. Third, the number of non-commuting observables needed to estimate molecular energies scales quartically $O(N^4)$ with the problem size~\cite{mcardle2020}, which could make quantum measurements highly impractical for larger molecules. Despite recent progress in developing strategies to reduce the number of measurements~\cite{hugginsEfficientNoiseResilient2021, kubler2020} and classical parameters~\cite{grimsley2019, tangQubitADAPTVQEAdaptiveAlgorithm2021}, VQA implementations for calculating molecular energy suffer from quantum noise, limiting quantum circuits to at most linear depth.

A fundamental starting point to determine the wavefunction and energy of a quantum many-body molecular system is the Hartree-Fock (HF) method, in which the exact $N$-body wavefunction of the system is approximated by a single determinant of one-electron orbital wavefunctions or molecular orbitals~\cite{szaboModernQuantumChemistry1982}. Recently, a linear depth quantum circuit ansatz with quadratic parameter count was implemented and demonstrated to evaluate HF energies of hydrogen chains and the barrier for diazene cis-trans isomerization on NISQ hardware~\cite{aruteHartreeFockSuperconductingQubit2020}. 

Although the HF approach provides about 99\% of the total molecular energy~\cite{helgakerCalibrationElectronicStructureModels2000}, it falls short in describing some key chemical phenomena. For instance, HF predicts that noble gas atoms do not attract each other at any temperature, and thus should not be able to liquefy. This is due to the treatment of electron-electron interactions in HF, where each electron experiences an average potential or 'mean-field' generated from all other electrons, with the instantaneous Coulomb interaction between any two electrons not properly accounted for. Such correlated motions of electrons, or electron correlation, have been attributed to be the largest source of error in quantum chemical calculations~\cite{martinElectronCorrelationNature2022}.

A number of approaches have been developed to accurately compute the energy that arises from electron correlation, which are usually carried out after a HF calculation, thus they are termed post-HF methods. The most straightforward is to consider the exact wavefunction being a linear combination of all possible excited electron configurations that can be generated from the HF wavefunction -- the full configuration interaction (FCI). However, the factorial scaling of FCI with number of electrons and basis functions makes it computationally costly except for very small molecular systems. Coupled cluster theory is another approach that converges to FCI with increasing numbers of excitations~\cite{bartlettCoupledClusterTheoryQuantum2007}. Notably, the unitary coupled cluster with single and double excitations (UCCSD) is a frequently encountered quantum computing ansatz~\cite{anandQuantumComputingView2022}, which scales as the sixth power with system size on classical computers. The UCCSD approach provides a fermionic unitary transformation that is easily translated into quantum circuits via standard fermion-to-qubit mappings such as the Jordan-Wigner~\cite{jordan1928} and Bravyi-Kitaev~\cite{bravyi2002, seeley2012}. However, despite best efforts in reducing quantum resources needed, the UCCSD ansatz requires at least $O(N^{3\sim5})$ circuit depth and $O(N^{2\sim4})$ number of parameters, which make it difficult to implement on currently available NISQ devices~\cite{anandQuantumComputingView2022}.

The approach used in our work is to apply many-body perturbation theory to the electronic Hamiltonian using parameterized HF wavefunctions as a starting point. That is, the electronic Hamiltonian is to be partitioned into an exactly solvable zeroth-order unperturbed term and a perturbative term as first proposed by M{\o}ller and Plesset~\cite{moller1934, cremer2011, popleTheoreticalModelsIncorporating1976}. The sum of the zeroth, first and second-order terms in the perturbative energy expansion yields the second-order M{\o}ller-Plesset (MP2) energy. In this work we demonstrate a variational NISQ-friendly algorithm for estimating the ground state energy based on minimizing the MP2 energy via optimizing the molecular orbitals, also known as orbital-optimized MP2 (OMP2) in quantum chemistry parlance~\cite{bozkaya2011a, bozkaya2013}. It is worth noting that our algorithm uses linear-depth circuits to estimate electronic energies that are inclusive of contributions from electron correlation.

The following sections will detail how we incorporate several recent innovations to estimate the OMP2 energy with a NISQ algorithm, which we will term as NISQ-OMP2. We first express the correlation energy in the perturbative expansion of the ground state energy into simple expectation values of a perturbation term with respect to the HF states using a technique from Projective Quantum Eigensolver~\cite{stair2021}. Second, we efficiently generate the required excited HF states via a fermionic double excitation evolution of a HF reference state using linear-depth quantum circuits~\cite{yordanov2020}. Third, we perform a parameterized orbital basis transformation on the HF states for orbital-optimization using linear-depth circuits via a QR decomposition scheme~\cite{clementsOptimalDesignUniversal2016}. Fourth, we efficiently estimate the correlation energy using a basis rotation grouping scheme that reduces $O(N^4)$ number of Pauli measurements down to $O(N)$~\cite{hugginsEfficientNoiseResilient2021}. In addition, to further reduce the number of qubits and circuit depth needed to represent the mean-field HF ansatz, we applied quantum embedding on large molecules to isolate the relevant molecular orbitals of interest for the calculation of the OMP2 energy~\cite{rossmannekQuantumHFDFTembedding2021}.

We tested NISQ-OMP2 on multiple cloud-accessed quantum processors and simulated quantum backends with and without quantum noise models by computing the OMP2 energies of H$_2$, H$_3^{+}$, LiH and H$_4$. We found that not only does NISQ-OMP2 reproduce the electronic energies around equilibrium internuclear distances, but it also captures the electron correlation-induced energy shifts at longer distances, despite obtaining low state fidelities with current noisy quantum processors.

\begin{figure*}
\centering
\includegraphics[width=1\textwidth, page=1]{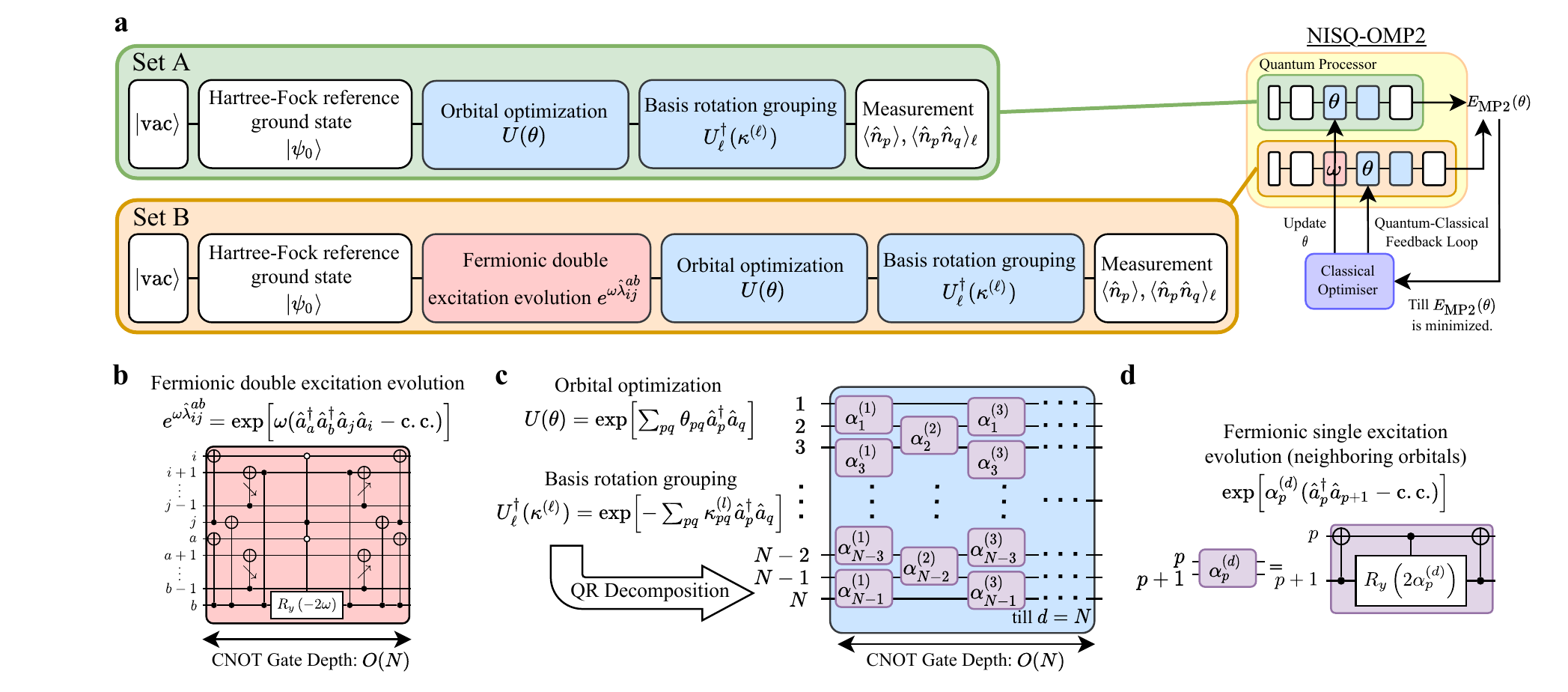}
\caption{(a) Quantum circuits sets in second quantization form and NISQ-OMP2 algorithm that minimizes MP2 energy $E_{\textrm{MP2}}\left(\theta\right)$ in Eq.~\eqref{oemp2} with respect to $\theta$. (b) Quantum gate decomposition of the fermionic double excitation evolution~\cite{yordanov2020} under the Jordan-Wigner (JW) Mapping. The CNOTs with diagonal pointing arrows denote CNOT ladders as given in Appendix~\ref{apx:gatedecomp}. (c) The orbital optimization and basis rotation grouping are QR decomposed~\cite{clementsOptimalDesignUniversal2016} and implemented separately on a quantum circuit, each with $N$ layers of parallel fermionic single excitation evolution (neighboring orbitals) which can be decomposed as shown in (d)~\cite{yordanov2020}.}
\label{fig1}
\end{figure*}

\section{Going Beyond HF in a NISQ Quantum Computer}
\subsection{Perturbing the Electronic Hamiltonian and Optimizing Orbitals}
We consider the general problem of finding the ground state energy of an electronic Hamiltonian $\hat{H}$ as given in the second-quantized form
\be
\hat{H} = \sum_{pq} h_{pq} \hat{a}_p^{\dagger} \hat{a}_q + \frac{1}{2} \sum_{pqrs} h_{pqrs} \hat{a}_p^{\dagger} \hat{a}_q^{\dagger} \hat{a}_r \hat{a}_s, \label{eq:H}\\
\ee
where $\hat{a}^{\dagger}_p$ and $\hat{a}_p$ are fermionic creation and annihilation operators for the $p$th molecular orbital, $h_{pq}$ are one-electron core integrals, and $h_{pqrs}$ are two-electron repulsion integrals. By solving the self-consistent field equations~\cite{szaboModernQuantumChemistry1982}, we obtain a mean-field Hamiltonian $\hat{F}$ which is a diagonal operator
\be
\hat{F} = \sum _{pq} \varepsilon_p\delta_{pq} \hat{a}_p^\dagger \hat{a}_q, \label{fock}\\
\ee
whose eigenstates are the molecular orbitals and eigenvalues $\varepsilon_p$ are the molecular orbital energies
\be
\varepsilon_p = h_{pp} + \sum_{i} \left(h_{piip} - h_{pipi}\right).\label{scf}
\ee
where we use indices $i$, $j$ to denote occupied orbitals.

In MP2, the electronic Hamiltonian $\hat{H} = \hat{F} + \hat{V}$ is partitioned into an unperturbed mean-field Hamiltonian $\hat{F}$ and a perturbation term $\hat{V}$~\cite{cremer2011, popleTheoreticalModelsIncorporating1976}. The perturbation expansion for the ground state energy up to second order, is simplified and grouped to give the MP2 energy $E_\mathrm{MP2}$ as follows
\begin{align}
E_\mathrm{MP2} =& \langle\psi_0|\hat{F}|\psi_0\rangle + \langle \psi_0|\hat{V}|\psi_0\rangle \nonumber \\
 & + \sum_{a<b, i<j} \frac{|\langle \psi_0 |\hat{V}|\psi_{ij}^{ab}\rangle|^2 }{\varepsilon_i+\varepsilon_j-\varepsilon_a-\varepsilon_b} \\
 =& E^{\left(0\right)}+E^{\left(1\right)} + E^{\left(2\right)}. \label{emp2}
\end{align}
where we use indices $a$, $b$ to denote virtual orbitals.

The zeroth order energy $E^{(0)}$ in Eq.~\eqref{emp2} is numerically calculated as the sum of all occupied orbital energies $\varepsilon_i$. The first order energy $E^{\left( 1 \right)}$ in Eq.~\eqref{emp2} is the mean-field electron interaction energy of the HF reference ground state $\left|\psi_0\right\rangle$. The sum of the zeroth and first order energies is the HF energy. The second order energy $E^{(2)}$ in Eq.~\eqref{emp2} is the first perturbative approximation of the electronic correlation energy, expressed in terms of off-diagonal elements of the perturbation $\hat{V}$. 

In OMP2 the total MP2 energy $E_\mathrm{MP2}$ is minimized via the optimization of the molecular orbitals. To facilitate such optimization we employ a parameterized unitary orbital transformation $\hat{U}(\theta)= \exp \left[ \sum_{pq} \theta_{pq} \hat{a}_p^\dagger \hat{a}_q \right]$ where $\theta$ is an anti-hermitian parameter matrix $\theta_{pq} = -\theta_{qp} (1-\delta_{pq})$. Applying this orbital transformation to the HF reference state and doubly-excited state, we obtain $|\psi_0(\theta)\rangle = \hat{U}(\theta)|\psi_0\rangle$ and $|\psi_{ij}^{ab}(\theta)\rangle = \hat{U}(\theta)|\psi_{ij}^{ab}\rangle$, respectively. After introducing this orbital transformation to the mean-field Hamiltonian $\hat{F}(\theta) = \hat{U}(\theta)\hat{F}\hat{U}^\dagger(\theta)$, the perturbation term can be written as $\hat{V}(\theta) = \hat{H}- \hat{F}(\theta)$. Inserting these parameters into Eq.~\eqref{emp2} yields the parameterized MP2 energy $E_\mathrm{MP2}\left(\theta\right)$,
\begin{align}
E_\mathrm{MP2}\left(\theta\right) =& \left\langle \psi_{0}(\theta)\right| \hat{F}(\theta)\left|\psi_{0}(\theta)\right\rangle +\left\langle \psi_{0}(\theta)\right|\hat{V}(\theta)\left|\psi_{0}(\theta)\right\rangle \nonumber \\
 & + \sum_{a<b,\:i<j}\frac{\left|\left\langle \psi_{0}(\theta)\left|\hat{V}(\theta) \right|\psi_{ij}^{ab}(\theta)\right\rangle \right|^{2}} {\varepsilon_{i}+\varepsilon_{j}-\varepsilon_{a}-\varepsilon_{b}} \\
 =& E^{\left(0\right)}\left(\theta\right) + E^{\left(1\right)}\left(\theta\right) + E^{\left(2\right)}\left(\theta\right). \label{oemp2}
\end{align}

The orbital transformation $\hat{U}(\theta)$ only mixes the eigenstates of mean-field Hamiltonian $\hat{F}$ into new ones, while the corresponding orbital energies $\varepsilon_{p}$ remain unaffected. $E^{\left( 0 \right)}$ in Eq.~\eqref{oemp2} is thus independent of $\theta$ and it is still equal to the sum of all occupied orbital energies $\varepsilon_i$.

For a general many-body system a truncated parameterized perturbation energy expansion is usually not variational, as the parameterization scheme can be arbitrary, which may result in an unbounded energy expansion in the parameter domain~\cite{kim2002, stevensonOptimizedPerturbationTheory1981}. Thus, it is typical to optimize the parameterized MP2 energy $E_{\textrm{MP2}}\left(\theta\right)$ based on the principle of least sensitivity~\cite{okopinskaNonstandardExpansionTechniques1987, stevensonOptimizedPerturbationTheory1981}, where the gradient magnitude $\left| \partial E_{\textrm{MP2}}\left(\theta\right)/\partial\theta \right|$ is minimized. However, due to the use of unitary orbital basis transformation $\hat{U}(\theta)$, all three perturbation terms $E^{\left( 0 \right)}\left(\theta\right)$, $E^{\left( 1 \right)}\left(\theta\right)$ and $E^{\left( 2 \right)}\left(\theta\right)$ are bounded in the $\theta$ domain. Therefore, the parameterized MP2 energy $E_{\textrm{MP2}}\left(\theta\right)$ is variational in $\theta$ and OMP2 energy $E_{\textrm{OMP2}}=\min_\theta \left[E_{\textrm{MP2}}\left(\theta\right)\right]$ may be obtained by variationally minimizing parameterized MP2 energy $E_{\textrm{MP2}}\left(\theta\right)$ with respect to $\theta$.

\subsection{Estimating OMP2 Energy Using Linear Depth Quantum Circuits}
Evaluating the off-diagonal elements of $\hat{V}(\theta)$ which determine the second order energy $E^{(2)}(\theta)$ poses a problem for NISQ devices as it apparently requires the use of a modified Quantum Hadamard test~\cite{aharonovPolynomialQuantumAlgorithm2009} or Swap test~\cite{barencoStabilizationQuantumComputations1997}, both of which require additional ancillary qubits and deep quantum circuits. Moreover, a direct quantum circuit implementation of the perturbation term $\hat{V}(\theta)$ under the JW mapping from fermions to qubits will result in $O(N^4)$ Pauli strings of quantum measurements, which is impractical beyond the smallest molecules~\cite{verteletskyiMeasurementOptimizationVariational2020}. We tackled both of these issues by incorporating the most effective solutions from recent developments. We build upon the effective measurement of residual elements in the projective quantum eigensolver approach~\cite{stair2021} to evaluate the off-diagonal elements of $\hat{V}(\theta)$. We have also applied the Basis-Rotating Grouping technique~\cite{pedersenDensityFittingAuxiliary2009, hugginsEfficientNoiseResilient2021} which performs a low-rank tensor factorization decomposition on the perturbation term $\hat{V}$ that reduces the number of Pauli string quantum measurements down to $O(N)$. In addition, we post-selected these quantum measurement results by the number of electrons to mitigate the readout measurement errors of quantum device, as the number of electrons is a conserved quantity in the entire quantum computation.

We let the off-diagonal elements be $r^{ab}_{ij}\left(\theta\right) = \langle \psi_0(\theta)|\hat{V}(\theta)|\psi_{ij}^{ab}(\theta)\rangle$ and define an anti-hermitian double excitation operator $\hat{\lambda}_{ij}^{ab} = \hat{a}^\dagger_a \hat{a}^\dagger_b \hat{a}_j \hat{a}_i - \textrm{c.c.}$. We apply an unitary double excitation operator with a fixed angle $\omega$, $e^{\omega \hat{\lambda}_{ij}^{ab}}$ on a HF reference state $|\psi_0\rangle$ as such
\be
e^{\omega\hat{\lambda}_{ij}^{ab}}|\psi_0\rangle = \cos \omega|\psi_0\rangle + \sin \omega |\psi_{ij}^{ab}\rangle,\label{lincombrefdoub}
\ee
Thus, the off-diagonal elements $r^{ab}_{ij}\left(\theta\right)$ can be evaluated in terms of ordinary expectation values and the first order energy $E^{\left( 1 \right)}(\theta)$ as follows
\be
r^{ab}_{ij}\left(\theta\right) = E^{ab,(1)}_{ij}\left(\theta\right) - \frac12 E^{ab}_{ij}\left(\theta\right) - \frac12 E^{(1)}(\theta),\label{residualrmu}
\ee
where
\begin{align}
E^{ab,(1)}_{ij} \left(\theta\right) &= \langle \psi_0|e^{-\frac{\pi}{4}\hat{\lambda}_{ij}^{ab}}\hat{U}{^\dagger}(\theta)\hat{V}(\theta)\hat{U}(\theta)e^{\frac{\pi}{4}\hat{\lambda}_{ij}^{ab}}|\psi_0\rangle, \nonumber\\
E^{ab}_{ij}\left(\theta\right) &= \langle \psi_{ij}^{ab}(\theta)|\hat{V}(\theta)|\psi_{ij}^{ab}(\theta)\rangle.
\end{align}

For an efficient quantum measurement of the perturbation term $\hat{V}(\theta)$, we applied a low-rank tensor factorization decomposition $\hat{V}(\theta)$ into $O(N)$ distinct summation terms as follows
\begin{align}
\hat{V} =& \hat{U}_0 \left(\kappa^{\left( 0 \right)}\right)  \left[ \sum _{p} d_{q} \hat{n}_p\right] \hat{U}_0^{\dagger}\left(\kappa^{\left( 0 \right)}\right) \nonumber \\ 
& + \sum _{l=1}^{O(N)} \hat{U}_l\left(\kappa^{\left( l \right)}\right)  \left[ \sum _{pq} d_{pq}^{\left( l \right)} \hat{n}_p \hat{n}_q   \right] \hat{U}_l^{\dagger}\left(\kappa^{\left( l \right)}\right),\label{basisrotobs}
\end{align}
where we dropped $\theta$ to reduce verbosity. Each summation term in Eq.~\eqref{basisrotobs} is a linear combination of orbital-transformed number operators $\hat{U}_l\left(\kappa^{\left( l \right)}\right)\hat{n}_p \hat{U}_l^\dagger\left(\kappa^{\left( l \right)}\right)$ and $\hat{U}_l\left(\kappa^{\left( l \right)}\right)\hat{n}_p\hat{n}_q\hat{U}_l^\dagger\left(\kappa^{\left( l \right)}\right)$ where $\hat{n}_p=\hat{a}_p^{\dagger}\hat{a}_p$ are number operators and $\hat{U}_l\left(\kappa^{\left( l \right)}\right) = \exp \left[ \sum_{pq} \kappa_{pq}^{\left( l \right)} \hat{a}_p^\dagger \hat{a}_q \right]$ are orbital transformations described by anti-hermitian matrices $\kappa^{\left( l \right)}$. Under a JW mapping, the number operators becomes a linear combination of Pauli-Z strings as $\hat{n}_p=\left(1+\hat{Z}_p\right)/2$. Thus, all number operators within each summation term in Eq.~\eqref{basisrotobs} can be simultaneously measured on a quantum computer. Therefore, we effectively have $O(N)$ quantum measurements. In actual implementation, we used Google's OpenFermion library~\cite{mccleanOpenFermionElectronicStructure2019} to perform low-rank tensor factorization decomposition on $\hat{V}\left(\theta\right)$ to get linear coefficients $d_{p}$, $d_{pq}^{\left( l \right)}$ and the basis rotations $\hat{U}^{\dagger}_{l}\left(\kappa^{\left( l \right)}\right)$ in Eq.~\eqref{basisrotobs}.

We have constructed two collections of quantum circuits, set A and B, to calculate $E_{\textrm{OMP2}}$ as shown in Fig.~\ref{fig1}(a). Quantum circuits in set A directly estimate the first order $E^{\left( 1 \right)}(\theta)$ in Eq.~\eqref{oemp2} by preparing the orbital-transformed HF reference ground state $\left|\psi_{g}(\theta)\right\rangle$. Meanwhile the quantum circuits in set B directly estimate the expectation values $E^{ab,(1)}_{ij} \left(\theta\right)$ and $E^{ab}_{ij} \left(\theta\right)$ by preparing quantum states $\hat{U}(\theta) e^{\frac{\pi}{4} \hat{\lambda}_{ij}^{ab}} |\psi_0\rangle$ and $|\psi_{ij}^{ab}(\theta)\rangle= \hat{U}(\theta) e^{\frac{\pi}{2} \hat{\lambda}_{ij}^{ab}} |\psi_0\rangle$, respectively. These two sets of quantum circuits will be executed on a quantum processor and the measurement results are used to estimate the parameterized MP2 energy $E_{\textrm{MP2}}\left(\theta\right)$, which is then fed into a classical optimizer that will generate a new set of parameters $\theta$ for the next optimization iteration. This iteration cycle creates a quantum-classical feedback loop which will eventually terminate upon the minimization of $E_{\textrm{MP2}}\left(\theta\right)$ up to a predetermined level of tolerance so as to output $E_{\textrm{OMP2}}$.

The quantum circuits for both the orbital optimization $\hat{U}(\theta)$ and basis rotation grouping $\hat{U}^{\dagger}_{l}\left(\kappa^{\left( l \right)}\right)$ can be optimally implemented with linear depth $O(N)$ via the optimal QR decomposition scheme~\cite{clementsOptimalDesignUniversal2016}, which is shallower by a constant factor than other recent approaches~\cite{kivlichanQuantumSimulationElectronic2018, aruteHartreeFockSuperconductingQubit2020}. This scheme involves implementing layers of fermionic single neighboring excitation evolution $\exp\left[\alpha_{p}^{\left(d\right)} \left( \hat{a}_p^\dagger \hat{a}_{p+1} - c.c.\right) \right]$, as shown in Fig.~\ref{fig1}(c). We detail the QR decomposition scheme in Appendix~\ref{apx:qrdecomp}. The quantum gate decomposition for the fermionic single neighboring excitation $\exp\left[\alpha_{p}^{\left(d\right)} \left( \hat{a}_p^\dagger \hat{a}_{p+1} - c.c.\right) \right]$ and double excitation $e^{\omega\lambda^{ab}_{ij}}$ evolutions, whose circuit depth scales linearly $O(N)$ in the number of qubits, are shown in Figs.~\ref{fig1}(d) and (b) respectively~\cite{yordanov2020}.

\subsection{Results and Benchmarking on NISQ Hardware on the Cloud}

\begin{figure}
\centering
\includegraphics[width=\columnwidth, page=1,]{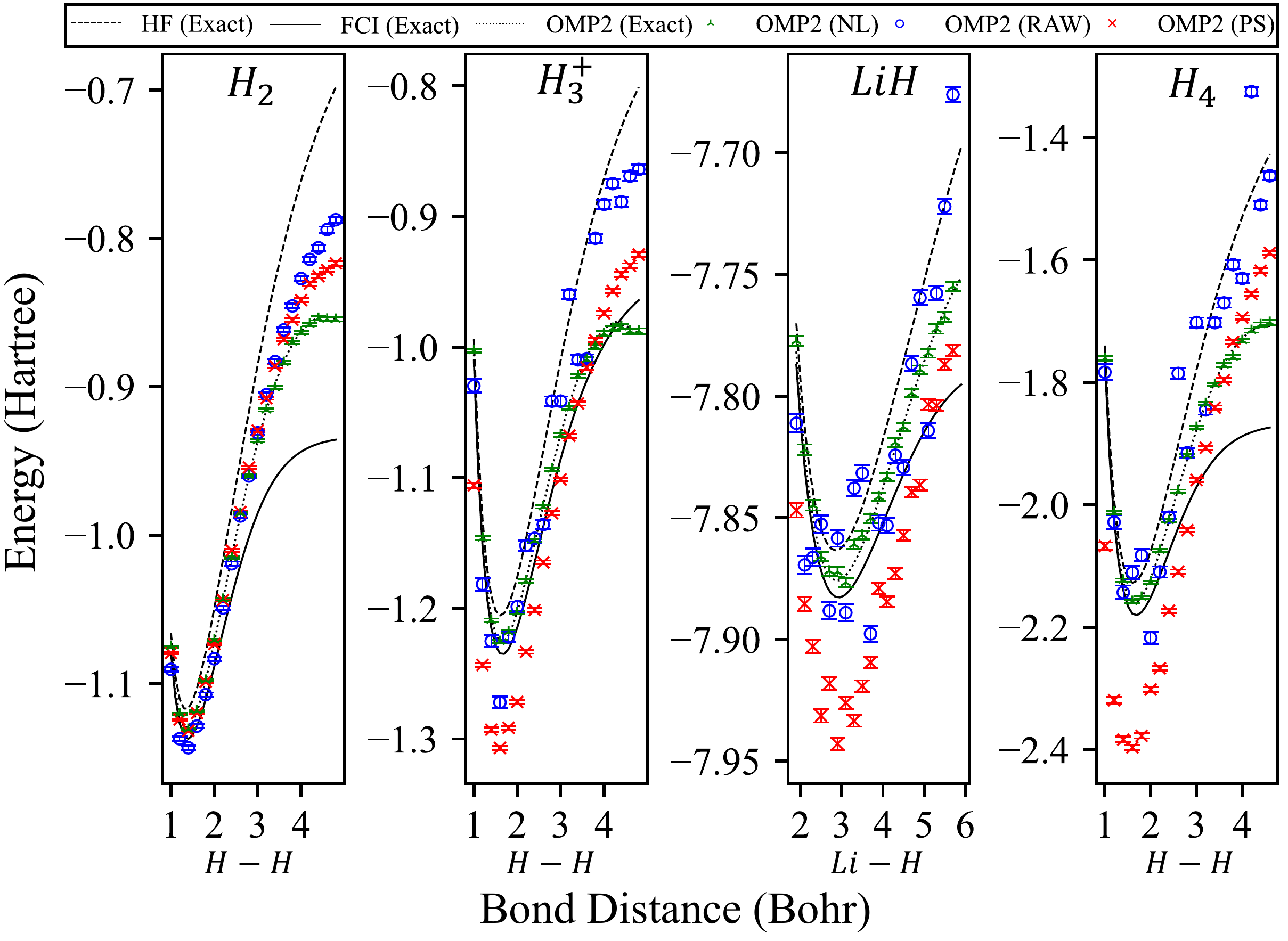}
\caption{Orbital-optimized MP2 energy $E_\mathrm{OMP2}$ for H$_2$, linear H$_3^{+}$, quantum-embedded LiH and linear H$_4$ at various bond distances obtained using a noiseless statevector simulator (green tri-ups) and the IBM-Qiskit simulated noise model of 27-qubit IBM Auckland with (red crosses) and without (blue circles) post-selection. The exact energies for HF, OMP2 and FCI calculated classically using Psi4 are plotted as dashed, dotted and solid black lines, respectively. The error bars are calculated from estimating the covariance matrix of the commuting Pauli-Z strings of the quantum measurements of the perturbation term $\hat{V}$ based on 100,000 shots per circuit.}
\label{fig2}
\end{figure}

The NISQ-OMP2 algorithm was tested on four molecules: (H$_2$, linear H$_3^{+}$, LiH and linear H$_4$) using the STO-3G basis set. This gives the number of spin orbitals, ($N=4$, $6$, $8$, $12$) and number of electrons ($N_e =2$, $2$, $4$, $4$) electrons respectively. We obtain the corresponding electronic Hamiltonian $\hat{H}$ by generating the one- and two-electron integrals using the Psi4 program~\cite{smithPSI4OpensourceSoftware2020}. Under the JW mapping, these molecular cases will require ($N=4$, $6$, $8$, $12$) qubits, respectively. We reduce the number of qubits required for the LiH from $N=12$ to $N=6$ using a quantum embedding technique~\cite{rossmannekQuantumHFDFTembedding2021}. An active space was created by partitioning 6 inactive spin-orbitals consisting of 2 occupied core spin-orbitals and 4 unoccupied excited spin-orbitals out of the total 12. The inactive spin-orbitals have a small overlap with 6 other active spin-orbitals that are close to the Fermi level. Thus, the quantum-embedded LiH molecule has $N_e=2$ active electrons with $N=6$ active spin-orbitals, making its problem size identical to the linear H$_3^{+}$ case. We define our orbital parameterization scheme of $\theta$ in Appendix~\ref{apx:varpara}, which involves $O(N^2)$ variational parameters to be optimized.

To show the best possible result achievable using the NISQ-OMP2 method, we pre-optimized the variational parameters $\theta$ on a quantum statevector simulator with a classical quasi-Newton L-BFGS-B optimizer. We found that the optimal parameter matrix for the hydrogen molecule H$_2$ case is an all-zero matrix, while the optimal parameters for the linear H$_3^{+}$, quantum-embedded LiH and linear H$_4$ molecules are given in Appendix~\ref{apx:varpara}. We implemented our optimized NISQ-OMP2 on an IBM-Qiskit~\cite{aleksandrowiczQiskitOpensourceFramework2019} simulated noise model of IBM-Auckland, several quantum processors: superconducting 27-qubit IBM-Auckland, superconducting 5-qubit IBM-Lima and ion trap 11-qubit IonQ~\cite{wrightBenchmarking11qubitQuantum2019}, all of which are accessed via quantum cloud providers: IBMQ and Amazon Braket. The noise parameters of these quantum processors are summarized in Appendix~\ref{apx:noiseinfo}.

We first classically simulate the NISQ-OMP2 circuits to understand the ideal performance of our approach. Figure~\ref{fig2} plots the OMP2 energy $E_{\mathrm{OMP2}}$ for various molecules (H$_2$, linear H$_3^{+}$, quantum-embedded LiH and linear H$_4$) as a function of the bond distance obtained using a noiseless statevector simulator and a simulated noise model for the IBM Auckland device with and without post-selection to mitigate errors. In the absence of noise (green triangles in Fig.~\ref{fig2}) we find good agreement with the exact OMP2 energy obtained numerically using Psi4, with small deviations arising due to shot noise from finite sampling (100k shots per distinct circuit). For all molecules, we observe that OMP2 energy is consistently well-below the HF energy, which shows that NISQ-OMP2 recovers the correlation energy around the equilibrium and intermediate bond distances as expected.

The raw OMP2 energies obtained under the simulated noise model are plotted as blue circles in Fig.~\ref{fig2}. Generally, the raw results have significant deviations from the corresponding exact OMP2 energies at longer bond distances, due to the second order energy $E^{(2)}$ becoming increasingly less accurate at longer bond distances, as shown in Appendix Fig.~\ref{fig:12_varmol}. As expected, we found that these deviations are caused by low quantum state fidelity of the deeper set B circuits, shown in Appendix Tab.~\ref{tab:statefidel}. These fidelities are improved by applying post-selection.

The corresponding post-selected OMP2 energies, shown in red crosses in Fig.~\ref{fig2}, were found to have fewer random errors than the raw OMP2 plots. This is likely due to the mitigation of the measurement readout errors as evidenced by the improved state fidelity in Appendix Tab.~\ref{tab:statefidel}. Despite this observed improvement, the state fidelity after post-selection remains significantly less than 90\%, which is very low, and is likely caused by large quantum gate errors present in current quantum processors. As a result, post-selected OMP2 energies were observed to be systematically less accurate than that of raw ones, suggesting the presence of fortuitous error cancellation.

\begin{figure}[H]
\centering
\includegraphics[width=0.95\columnwidth, page=2]{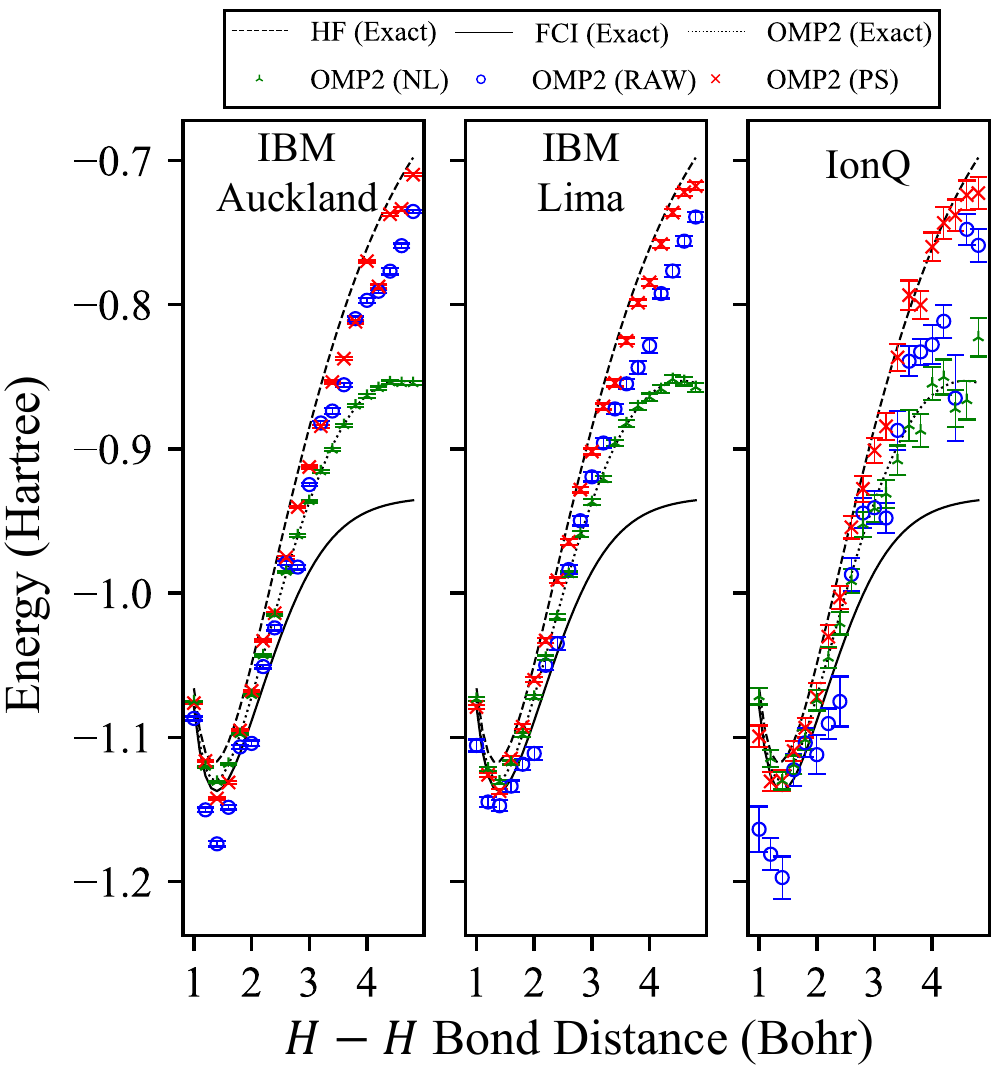}
\caption{Orbital-optimized MP2 energy $E_\mathrm{OMP2}$ for H$_2$ at various bond distances evaluated using different cloud quantum processors with (red crosses) and without (blue circles) post-selection: 27-qubit IBM Auckland (100,000 shots per circuit), 5-qubit IBM Lima (20,000 shots per circuit), and 11-qubit IonQ (1,000 shots per circuit). For comparison, the results of a noiseless statevector simulator are plotted in green triangles, and the exact energies for HF, OMP2 and FCI calculated classically using Psi4 are shown as dashed, dotted and solid black lines, respectively.}
\label{fig3}
\end{figure}

Next, we evaluated the OMP2 energy $E_\mathrm{OMP2}$ for the H$_2$ molecule on IBM Auckland (100k shots per circuit), IBM Lima (20k shots per circuit), and IonQ (1k shots per circuit). The raw energies are shown as blue circles in Fig.~\ref{fig3}. For all tested cloud quantum processors, the raw OMP2 energy plots show a clear qualitative description of electronic energy around equilibrium H$_2$ bond distance of 1.4 Bohr, with random errors descreasing with the number of shots used.

As expected, the corresponding post-selected OMP2 energy, as given in red crosses in Fig.~\ref{fig3}, was observed to have fewer random errors than the raw results. Generally, post-selection also led to more accurate energies around the equilibrium bond distance, but deviations occur for longer bond distances. As discussed above, this is mainly due to the lower quantum state fidelity of the deeper set B circuits (Appendix Tab.~\ref{tab:statefidel}). This generally causes the post-selected first order energy $E^{(1)}$ to be relatively more accurate than that of post-selected second order en-
\noindent ergy $E^{(2)}$, as shown in the Appendix Fig.~\ref{fig:12_h2mol}. As a result, post-selected OMP2 energy data exhibit a closer match to classically obtained values around the equilibrium bond distance where, the first order energy $E^{(1)}$ is most significant, while it deviates at longer bond distance where post-selected second order energy $E^{(2)}$ is becomes more important.

\begin{figure}
\centering
\includegraphics[width=\columnwidth]{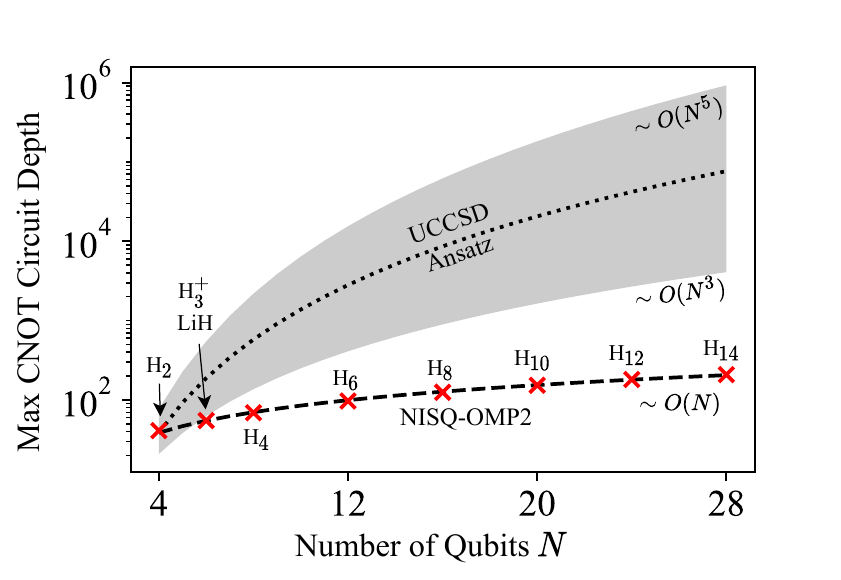}
\caption{Maximum CNOT gate depth required by NISQ-OMP2 and compared against UCCSD ansatz for various molecules in STO-3G atomic basis. Since we have used quantum embedding for the LiH molecule it only requires 6 qubits.}
\label{fig4}
\end{figure}

Finally, we compare the quantum resources required by NISQ-OMP2 algorithm against a state-of-the-art VQA that employs UCCSD ansatzes. A comparison of the required CNOT gate circuit depths between both algorithms is plotted in Fig.~\ref{fig4} for various molecules used in this study and larger hydrogen chains up to H$_{14}$ in the STO-3G atomic basis. For molecules (H$_2$, linear H$_3^{+}$, quantum-embedded LiH and linear H$_4$), the maximum CNOT circuit depth used in NISQ-OMP2 are estimated to be (41, 55, 55, and 69), respectively, which are generally shallower than that of UCCSD about ($\sim$40, $\sim$100, $\sim$100, $\sim$600). This comparison shows that the circuit depth required by NISQ-OMP2 scales linearly in $N$ number of qubits, which is a polynomial improvement of $O(N^{2\sim4})$ compared to the UCCSD approach, in exchange for $O(N^4)$ times more quantum circuit evaluations under Basis Rotation Grouping. Thus, while NISQ-OMP2 requires more circuit evaluations, we anticipate it will be able to handle larger systems than currently feasible using UCCSD due to the exponential loss of fidelity for deeper circuits. A detailed breakdown of the quantum resource estimates for molecules used in this paper is given the Appendix Tab.~\ref{tab1}.

\section{Discussion and Conclusion} 

The NISQ-OMP2 algorithm provides an estimation of ground state electronic energies that include explicit electron correlation effects using linear depth quantum circuits, and an improved resilience to noise compared to other variational quantum algorithms requiring deeper circuits. As the fidelities of quantum processors improve, it is expected that NISQ-OMP2 will enable the study of moderately larger molecular systems without requiring full quantum error correction.

Applications of NISQ-OMP2 can easily be extended beyond estimating ground state electronic energies to other quantum chemistry problems where M{\o}ller-Plesset perturbation theory has been fruitfully applied. These include the estimation of atomization energies, electron affinities, and ionization potentials of covalent systems, and other quantities that are difficult to obtain by experiment, such as the interaction energy between noble gas atoms.

To realize these potential applications, further improvements NISQ-OMP2 algorithm can be introduced improve its practicality and accuracy. For example, instead of relying on the quantum cloud providers' circuit transpilers, which may be not depth-optimal, direct transpilation of the quantum circuit may be used to leverage NISQ hardware's native quantum gates to reduce the overall quantum circuit depth and improve the quantum fidelity of fermionic excitation evolutions. NISQ-OMP2 can be sped up by executing multiple quantum circuits from set A and B in parallel on the same quantum device to reduce the total number of quantum circuit repetitions, in exchange for requiring more qubits. It will also be interesting to explore whether other fermion-to-qubit mappings such as Bravyi-Kitaev~\cite{bravyi2002, seeley2012} can be used to further reduce the number of qubits and circuit depth.

Another important direction for future work is to carry out the orbital optimization on real NISQ devices using noise-resilient classical optimization techniques such as simultaneous perturbation stochastic approximation~\cite{spallImplementationSimultaneousPerturbation1998}, particle swarm optimization~\cite{bonyadiParticleSwarmOptimization2017}, etc. In addition, other forms of error-mitigation techniques such as zero-noise extrapolation~\cite{liEfficientVariationalQuantum2017, temmeErrorMitigationShortDepth2017a, kandalaExtendingComputationalReach2019} may help to improve the state fidelity and provide a more accurate energy estimate.

However, we observed effects of error cancellation where mitigating the quantum measurement readout error alone may not always yield better results, despite the improved state fidelity after post-selection. In some cases these mitigation measures may even deviate the obtained results further. This highlights the importance of reducing quantum gate errors to further increase the state fidelity, which unfortunately remains an difficult issue in the NISQ era. Current quantum error correction (QEC) techniques are infeasible to implement as it requires many ancillary qubits of multiple factors~\cite{gottesmanIntroductionQuantumError2009}. 

In conclusion, we have proposed and implemented a shallow NISQ-friendly algorithm that estimates the electronic energies, including those from electron correlation effects, of molecular systems using linear-depth $O(N)$ quantum circuits via perturbation theory. Although such an approach is not superior to the UCCSD approach -- which requires at least $O(N^{3\sim5})$ circuit depth -- in the recovery of electron correlation, NISQ-OMP2 avoids exponentially lower state fidelities that result from deep circuits. To the best of our knowledge, this is the first linear-depth quantum algorithm that provides an estimate for correlation energy perturbatively using minimal heuristics.

\section*{Acknowledgements}
We acknowledge support from the National Research Foundation, Prime Minister’s Office, Singapore under the Quantum Engineering Programme (NRF2021-QEP2-02-P02) and the Agency for Science, Technology and Research (\#21709). We thank IBM and AWS for cloud quantum computer access.

\bibliographystyle{plainnat}
\bibliography{references.bib}

\clearpage
\onecolumn
\newpage
\appendix
\section{Quantum Gate Decomposition of Circuits} \label{apx:gatedecomp}

Under a JW mapping, the fermionic single neighboring excitation evolution $\exp\left[\alpha_{p}^{\left(d\right)} \left( \hat{a}_p^\dagger \hat{a}_{p+1} - c.c.\right) \right]$ can be decomposed in $3$ CNOT circuit depth as shown in Fig.~\ref{fig1}(d). Thus, the CNOT circuit depth of every quantum circuit in set A is exactly $6N$.

The fermionic double excitation evolution $e^{\omega \hat{\lambda}_{ij}^{ab}}$ can be decomposed using 2 parallel pairs of CNOT ladders and a multi-controlled Rotation Pauli-Y~\cite{yordanov2020}, as shown in Fig.~\ref{fig1}(b), which is an 8-fold CNOT depth reduction over the traditional 16 sequential CNOT ladders~\cite{whitfield2011}. We decomposed the multi-controlled Rotation Pauli-Y in the double excitation evolution using the standard technique \cite{barenco1995} that uses 13 CNOT circuit depth. As a result, the fermionic double excitation evolutions has a CNOT gate depth of $17+2\left( 1-\delta_{j,i+1}\right)+2\textrm{max}\left\{0,j-i-2, b-a-2\right\}$, assuming all-to-all qubit connectivity. Therefore, the maximum CNOT gate depth in set B is $O(N) \approx 6N+17+2\left( 1-\delta_{N_e,2}\right)+2\textrm{max}\left\{0,N_e-3, N-N_e-3\right\}$ that corresponds to the double excitation index $i=1$, $j=N_e$, $a=N_e+1$, $b=N$. 

We compare the CNOT circuit depth required for NISQ-OMP2 against the 1st Order Trotterized-UCCSD approach~\cite{anandQuantumComputingView2022} in Tab.~\ref{tab1}. It has recently come to our attention during the preparation of the paper that there already exists more CNOT-efficient gate decompositions of multi-controlled rotation Pauli-Y requiring 5 fewer CNOTs~\cite{yordanov2020}, which we did not include in our circuits and in our gate counts.

\begin{figure}[H]
\centering
\includegraphics[width=0.9\columnwidth, page=1]{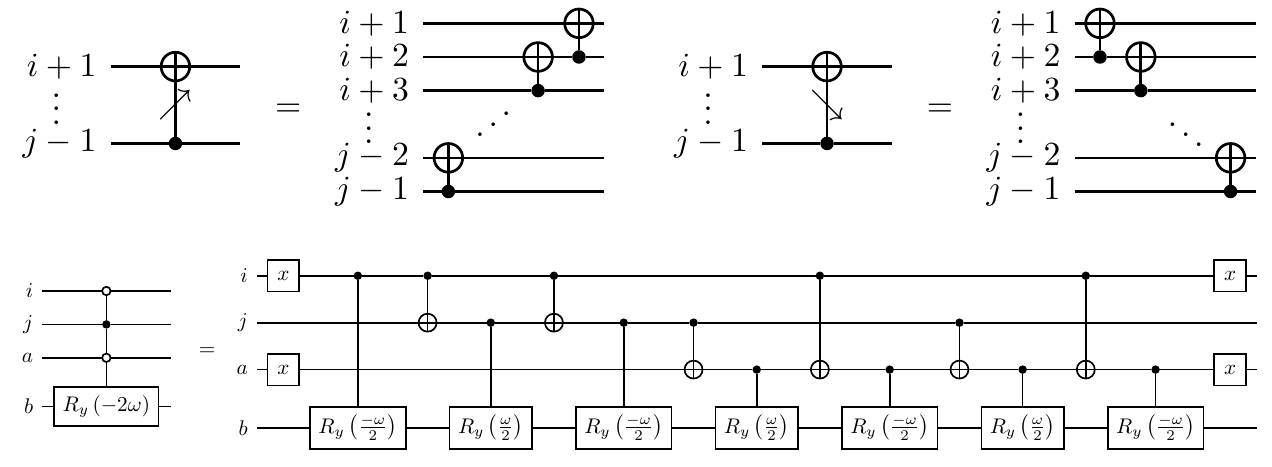}
\caption{Representation of CNOT ladders and explicit quantum circuit decomposition of multi-controlled rotation Pauli-y gate~\cite{barenco1995}.}
\label{fig:gatedecomp}
\end{figure}

\begin{table}[H]
\centering
\begin{tabular}{|l|c|c|c|c|c|}
\hline
NISQ-OMP2 & Scaling & H$_2$ & H$_3^+$ & LiH & H$_4$ \\ \hline
No. of Qubits & $N$ & 4 & 6 & 6 & 8 \\ \hline
No. of Variational Parameters & $O(N^2)$ & 1 & 2 & 2 & 4 \\ \hline
Max CNOT Depth in Set A & $O(N)$ & 24 & 36 & 36 & 48 \\ \hline
Max CNOT Depth in Set B & $O(N)$ & 41 & 55 & 55 & 69 \\ \hline
Total No. of Circuits (Set A \& B) & $O(N^5)$ & 12 & 91 & 91 & 803 \\ \hline\hline
UCCSD &~&~&~&~&~\\\hline
Max CNOT Depth for UCCSD & $O(N^5)$ & $\sim$40 & $\sim$100 & $\sim$100 & $\sim$600 \\\hline
\end{tabular}
\caption{Quantum resource estimates of NISQ-OMP2 against 1st Order Trotterised UCCSD approach (One Trotter Repetition), assuming all-to-all qubit connectivity on a quantum processor, for molecules used in the results of the paper.}
\label{tab1}
\end{table}

\newpage
\section{QR Decomposition of Orbital Optimization Quantum Circuit} \label{apx:qrdecomp}

We summarize the steps of the optimal QR decomposition of orbital optimization $U\left(\theta \right)$ ~\cite{clementsOptimalDesignUniversal2016} and derive the corresponding quantum circuit for any arbitrary orbital basis transformation $\hat{U}(\theta)= \exp \left[ \sum_{pq} \theta_{pq} \hat{a}_p^\dagger \hat{a}_q \right]$. The goal is to perform QR decomposition on $\hat{U}(\theta)$ into Eq.~\eqref{efficientoo} on the condition that the parameter matrix is anti-hermitian $\theta_{pq} = -\theta_{qp} (1-\delta_{pq})$, so that it can be implemented as $N$-layers of parallel of fermionic single neighboring excitations evolution $\hat{S}_{p}\left(\alpha_{p}^{\left(d\right)} \right)=\exp\left[\alpha_{p}^{\left(d\right)} \left( \hat{a}_p^\dagger \hat{a}_{p+1} - c.c.\right) \right]$. The following are a series of steps to achieve this objective. 

\begin{figure}[H]
\centering
\includegraphics[width=0.25\columnwidth, page=3]{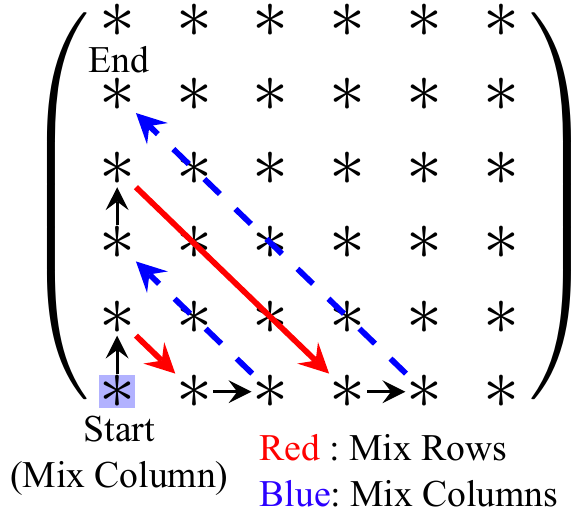}
\caption{Zig-zag route of zeroing elements in matrix form of orbital basis transformation $U$ in $\hat{a}_p^\dagger \hat{a}_q$ basis under the optimal QR decomposition scheme.}
\label{fig:zig-zag}
\end{figure}

First, the orbital basis transformation $\hat{U}$ is expressed in a 2D matrix form $U$ in the $\hat{a}_p^\dagger \hat{a}_q$ basis. We recognize that applying $\hat{S}\left(\alpha_{p}^{\left(d\right)}\right)$ on the left of $\hat{U}$ is effectively a Givens rotation in the $\hat{a}_p^\dagger \hat{a}_q$ basis that mixes rows $p$ and $p+1$ of $U$ and similarly applying $\hat{S}^\dagger\left(\alpha_{p}^{\left(d\right)}\right)$ on the right of $\hat{U}$ mixes columns $p$ and $p+1$ of $U$. Then, we may decompose $U$ as a series of Givens rotations by zeroing out lower triangle elements of $U$ in a zig-zag manner as shown in Fig.~\ref{fig:zig-zag}, alternating between mixing row and columns to zero out the corresponding elements. We start by zeroing out the bottom-right most element $U_{N,1}$ by mixing the corresponding columns with $\hat{S}_{1}^\dagger\left(\tan^{-1}\frac{U_{N,1}}{U_{N,2}}\right)$. Next, we move up to the next element $U_{N-1,1}$; note that this element has been mixed earlier and it is thus different from the initial value. We zero out this element $U_{N-1,1}$ by mixing the corresponding rows with $\hat{S}_{N-2}\left(\tan^{-1}\frac{U_{N-1,1}}{U_{N-2,1}}\right)$. We repeat the mixing of the row and column along the zig-zag path until all the lower triangle elements of $U$ are zero out. Eventually $U$ will become an identity, as our initial parameter matrix $\theta$ does not have any diagonal elements that correspond to phase shifts in $\hat{a}_p^\dagger \hat{a}_p$.

\be
\exp\left[\sum_{pq} \theta_{pq} \hat{a}_p^\dagger \hat{a}_q \right]=\prod_{d=odd}^{N}\left\{ \prod_{p=\textrm{odd}}^{N} \exp\left[\alpha_{p}^{\left(d\right)} \left(  \hat{a}_p^\dagger \hat{a}_{p+1} - c.c.\right) \right] \prod_{q=\textrm{even}}^{N} \exp\left[ \alpha_{q}^{\left(d+1\right)} \left(\hat{a}_q^\dagger \hat{a}_{q+1} - c.c. \right) \right] \right\} \label{efficientoo}
\ee

\section{Orbital Optimization} \label{apx:varpara}
\begin{figure}[H]
\centering
\includegraphics[width=0.5\columnwidth]{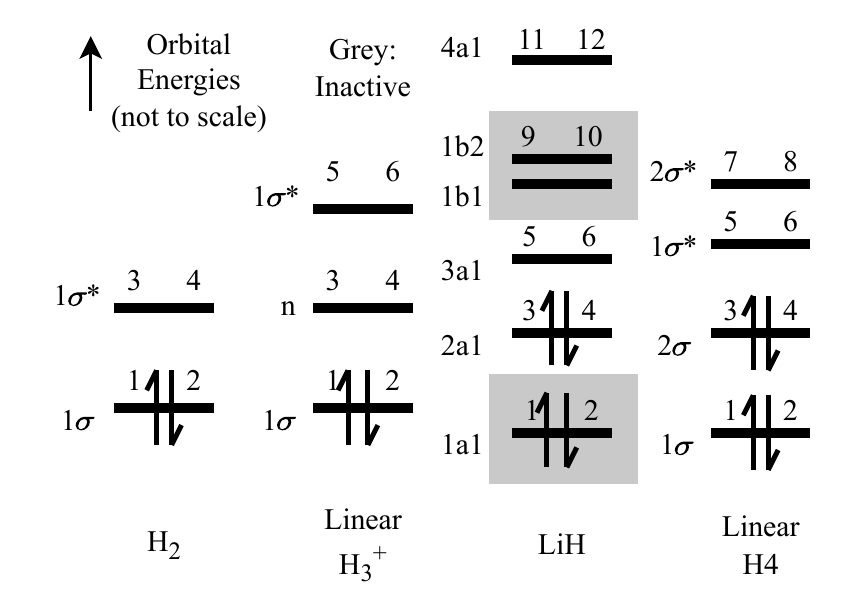}\hfill

\caption{Schematic HF molecular orbital (MO) diagrams for various molecules, using the STO-3G atomic orbital basis, with their symmetry labels on the left. $\sigma$, $\sigma$*, and $n$ orbitals are bonding, antibonding, and nonbonding MOs, respectively. For LiH, the symmetry labels for C$_{2 v}$ point group, which was utilized by the Psi4 code, are shown. Labelled in grey are inactive orbitals that are specified in the quantum embedding scheme for LiH, namely those for the core 1a1 MO, and virtual MOs that have differing symmetry with the 2a1 MO. The remaining active MOs for LiH are studied using the OMP2 method.}
\label{fig:orbitalenergies}
\end{figure}

A schematic diagram of the index-labelled HF molecular orbitals of the H$_2$, linear H$_3^+$, quantum-embedded LiH and linear H$_4$ is shown in Fig.~\ref{fig:orbitalenergies}. We assume the restricted HF case which enforces equal orbital transformation between both spins of same spatial orbitals, thus, we only consider mixing occupied and unoccupied spatial orbitals. As a result, the parameter matrix $\theta$ is a $N\times N$ matrix with $\frac{N_e\left( N-N_e \right)}{4} \approx O(N^2)$ number of unique variational parameters, such that $\theta_{pq}=-\theta_{qp}=\theta_{p+1,q+1}=-\theta_{q+1,p+1}$, $\forall\,p=\textrm{odd}\,i $ and $q=\textrm{odd}\,a$, and $\theta_{pq}=0$ if otherwise. The unique molecular orbital indices $p$, $q$ pairs are summarized in Tab. \ref{tab:optparams} (Top). For example, the parameter matrix $\theta$ of hydrogen molecule H$_2$ will be a $4\times4$ matrix and it will be described by one unique parameter such that $\theta_{13}=-\theta_{31}=\theta_{24}=-\theta_{42}$ and $\theta_{pq}=0$ if otherwise.

We initialize our variational parameters at the origin and implemented Scipy L-BFGS-B optimizer for orbital optimization. The average number of L-BFGS-B iterations for H$_2$, linear H$_3^+$, quantum-embedded LiH and linear H$_4$ are (1, 4, 4, 5), respectively. The L-BFGS-B optimized parameters for H$_2$ was found to be exactly zero for all bond distances while the others are given in Tab.~\ref{tab:optparams}

\begin{table}[H]
\begin{adjustbox}{width=\columnwidth,center}
\begin{tabular}{|c|c|c||c|c|c||c|c|c|c|c|}
\hline
\multicolumn{11}{|c|}{Optimized Parameters (Unique $(p,q)$ odd index)}  \\ \hline
\multicolumn{3}{|c||}{H$_3^+$} & \multicolumn{3}{c||}{Quantum-Embedded LiH} &  \multicolumn{5}{c|}{H$_4$}   \\ \hline
Bond (Bohr)  & (1,3)& (1,5)  & Bond (Bohr) &(3,5)  &(3,11) & Bond (Bohr) & (3,5)& (1,5)  &(3,7) & (1,7) \\ \hline
1.0 & 9.59E-09 & -3.27E-03 & 1.9 & -9.96E-03 & -1.00E-03 & 1.0 & -2.76E-08 & 3.03E-03 & -7.36E-04 & 7.95E-09  \\ \hline
1.2 & 1.19E-08 & -5.10E-03 & 2.1 & -1.08E-02 & -1.40E-03 & 1.2 & 3.48E-08 & 3.52E-03 & -1.24E-03 & 6.74E-08  \\ \hline
1.4 & 1.93E-08 & -7.48E-03 & 2.3 & -1.17E-02 & -1.83E-03 & 1.4 & 7.02E-08 & 3.66E-03 & -1.63E-03 & -5.25E-08  \\ \hline
1.6 & 1.70E-08 & -1.04E-02 & 2.5 & -1.26E-02 & 2.25E-03 & 1.6 & 1.67E-08 & 3.31E-03 & -1.74E-03 & 1.67E-08  \\ \hline
1.8 & 2.63E-08 & -1.40E-02 & 2.7 & -1.35E-02 & 2.64E-03 & 1.8 & 1.98E-08 & 2.33E-03 & -1.40E-03 & 1.79E-08  \\ \hline
2.0 & 7.46E-09 & -1.82E-02 & 2.9 & -1.47E-02 & 3.03E-03 & 2.0 & 2.21E-08 & 4.01E-04 & -5.14E-04 & 2.21E-08  \\ \hline
2.2 & 8.39E-09 & -2.32E-02 & 3.1 & -1.61E-02 & 3.41E-03 & 2.2 & 2.74E-08 & -2.15E-03 & 1.49E-03 & 1.37E-08  \\ \hline
2.4 & 3.70E-08 & -2.90E-02 & 3.3 & -1.78E-02 & 3.82E-03 & 2.4 & -4.74E-08 & -6.00E-03 & 4.66E-03 & -3.15E-09  \\ \hline
2.6 & 1.01E-08 & -3.56E-02 & 3.5 & -1.99E-02 & 4.28E-03 & 2.6 & 1.03E-07 & -1.14E-02 & 9.36E-03 & 5.81E-08  \\ \hline
2.8 & 2.17E-08 & -4.30E-02 & 3.7 & -2.25E-02 & 4.82E-03 & 2.8 & 8.91E-08 & -1.82E-02 & 1.56E-02 & -2.47E-08  \\ \hline
3.0 & 1.74E-08 & -5.11E-02 & 3.9 & -2.56E-02 & 5.51E-03 & 3.0 & -6.83E-08 & -2.64E-02 & 2.35E-02 & -4.07E-08  \\ \hline
3.2 & -1.22E-08 & -5.97E-02 & 4.1 & -2.93E-02 & 6.41E-03 & 3.2 & -6.75E-08 & -3.61E-02 & 3.30E-02 & 4.70E-08  \\ \hline
3.4 & 1.90E-08 & -6.88E-02 & 4.3 & -3.37E-02 & 7.62E-03 & 3.4 & 1.28E-08 & -4.58E-02 & 4.26E-02 & 1.84E-08  \\ \hline
3.6 & -6.56E-09 & -7.82E-02 & 4.5 & -3.87E-02 & 9.27E-03 & 3.6 & -2.25E-08 & -5.49E-02 & 5.19E-02 & 3.70E-08  \\ \hline
3.8 & -4.04E-08 & -8.78E-02 & 4.7 & -4.44E-02 & 1.15E-02 & 3.8 & -6.03E-08 & -6.28E-02 & 6.00E-02 & 4.08E-08  \\ \hline
4.0 & 1.55E-11 & -9.76E-02 & 4.9 & -5.10E-02 & 1.46E-02 & 4.0 & -1.12E-07 & -6.90E-02 & 6.65E-02 & 1.64E-07  \\ \hline
4.2 & 7.02E-09 & -1.07E-01 & 5.1 & -5.80E-02 & 1.86E-02 & 4.2 & -1.72E-08 & -7.33E-02 & 7.11E-02 & 4.34E-08  \\ \hline
4.4 & -3.58E-08 & -1.17E-01 & 5.3 & -6.54E-02 & 2.38E-02 & 4.4 & 5.76E-08 & -7.56E-02 & 7.37E-02 & -6.88E-08  \\ \hline
4.6 & 2.91E-08 & -1.27E-01 & 5.5 & -7.28E-02 & 3.03E-02 & 4.6 & 8.68E-09 & -7.62E-02 & 7.46E-02 & -4.94E-09  \\ \hline
4.8 & 6.88E-08 & -1.38E-01 & 5.7 & -7.96E-02 & 3.81E-02 & - & - & - & - & - \\ \hline
\end{tabular}
\end{adjustbox}
\caption{Unique L-BFGS-B optimized variational parameters elements as defined above for linear H$_3^+$, quantum-embedded LiH and linear H$_4$ at various bond distances.}
\label{tab:optparams}

\end{table}

\section{First and Second Order Energies} \label{apx:12energy}

\begin{figure}[H]
\centering
\includegraphics[width=0.48\columnwidth, page=4]{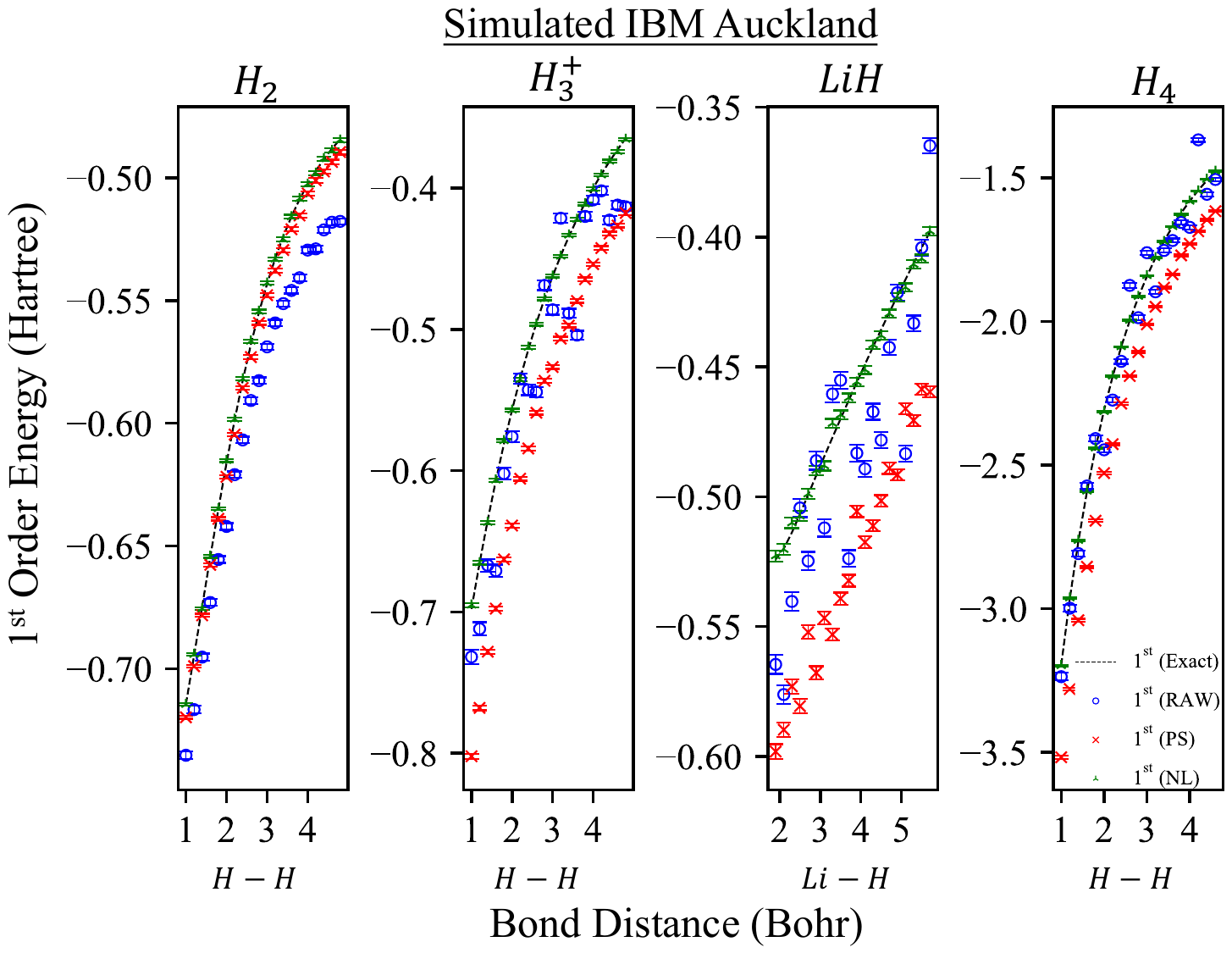}\hfill
\includegraphics[width=0.48\columnwidth, page=5]{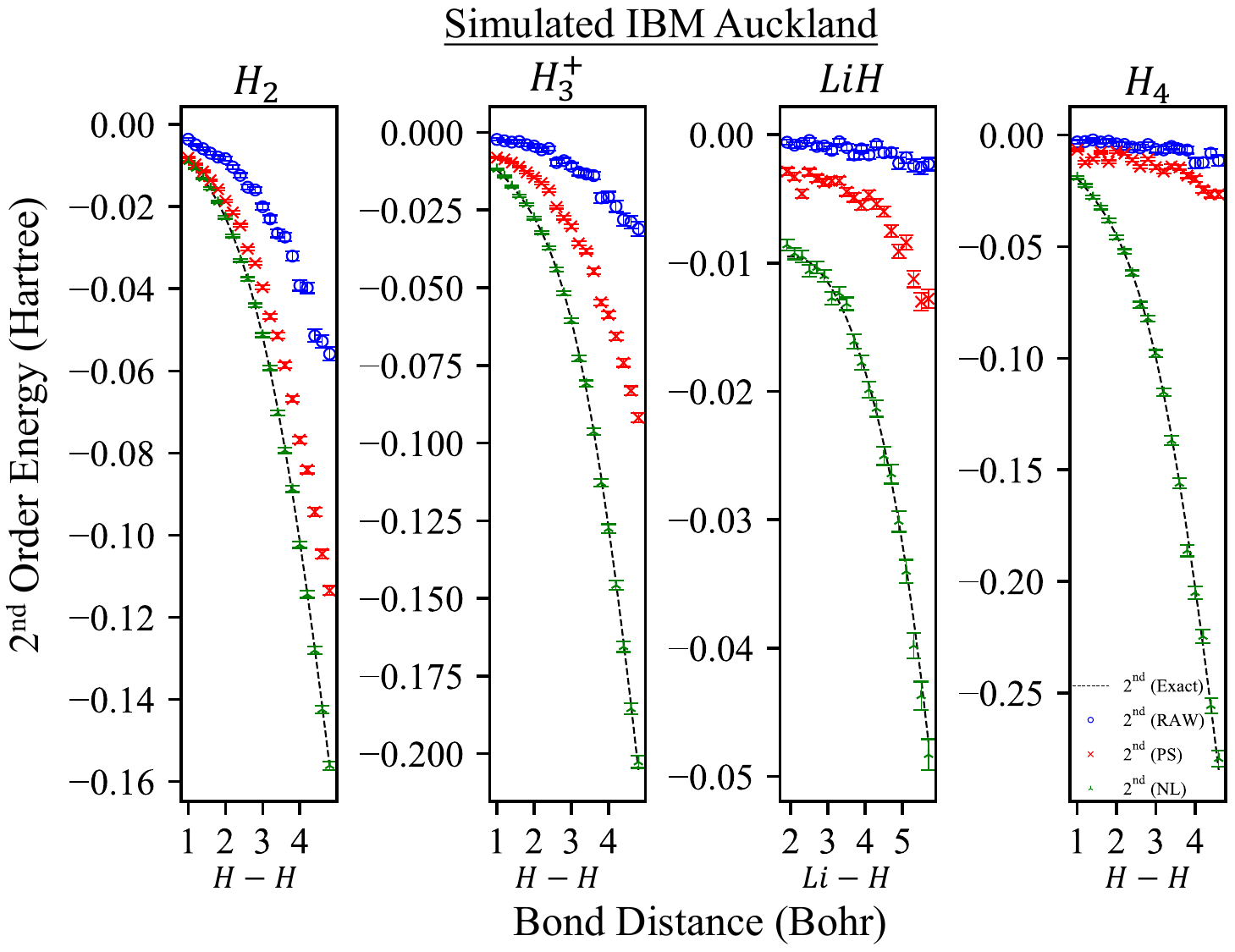}
\caption{First Order $E^{\left( 1 \right)}(\theta^*)$ and Second Order  $E^{\left( 2 \right)}(\theta^*)$ in Eq.~\eqref{oemp2}, where $\theta^*$ are the optimized parameters, for H$_2$, linear H$_3^+$, quantum-embedded LiH and linear H$_4$ for various bond distances, implemented on a Qiskit simulation of a noisy quantum processor with 100,000 shots per circuit. The noise model is obtained from Superconducting IBM Auckland.}
\label{fig:12_varmol}
\end{figure}

\begin{figure}[H]
\centering
\includegraphics[width=0.38\columnwidth, page=6]{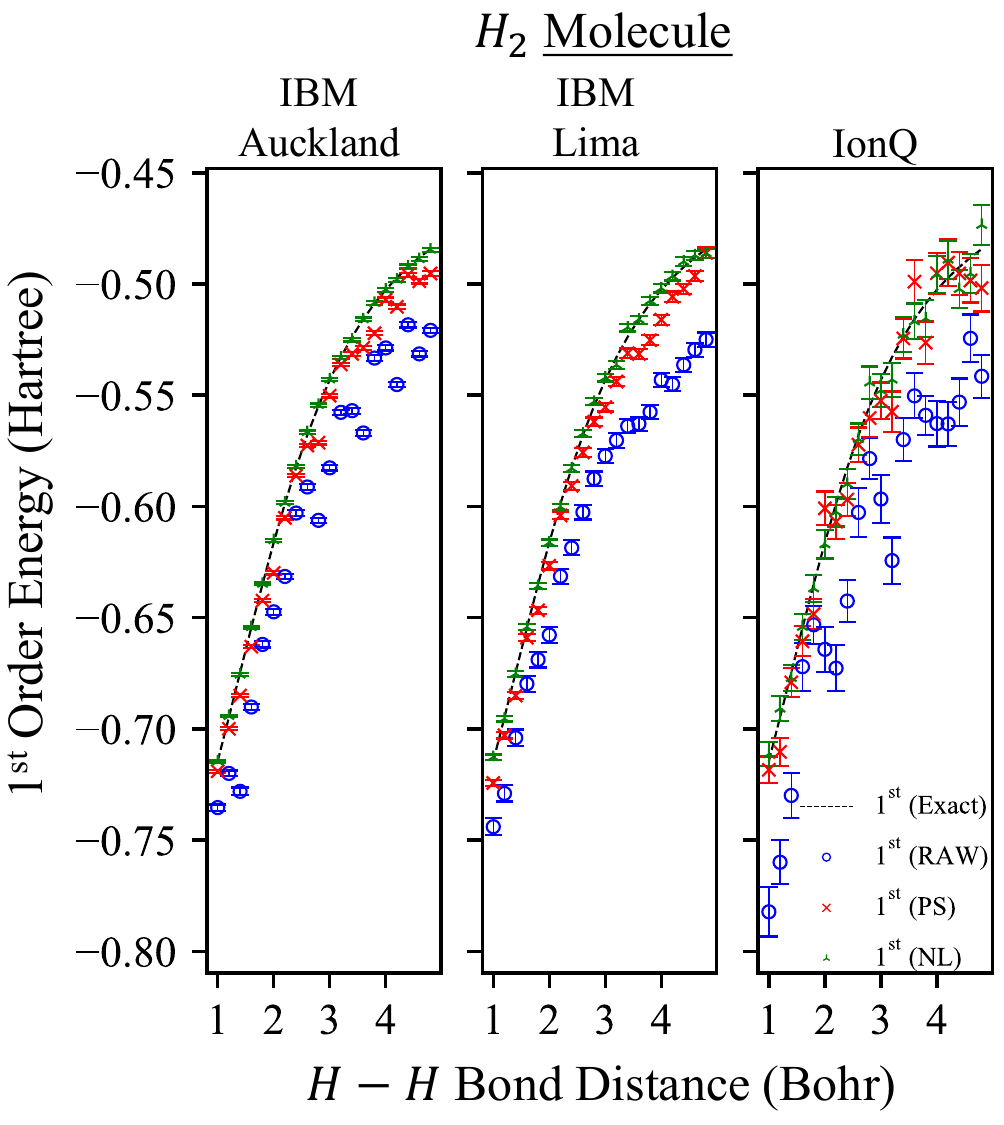}\hspace{0.05\textwidth}
\includegraphics[width=0.38\columnwidth, page=7]{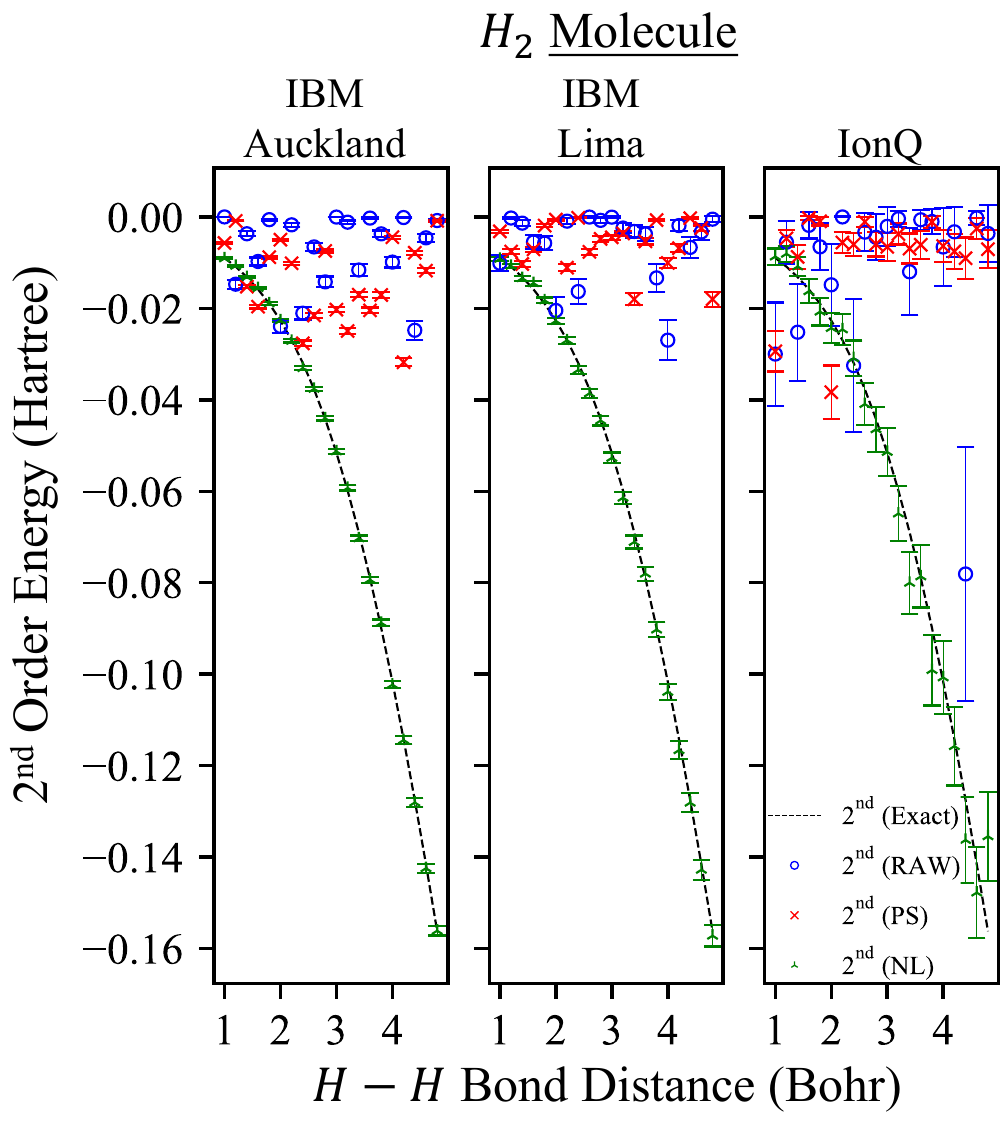}
\caption{First Order Energy $E^{\left( 1 \right)}(\theta^*)$ and Second Order  $E^{\left( 2 \right)}(\theta^*)$ Eq.~\eqref{oemp2}, where $\theta^*$ are the optimized parameters, for H$_2$ for various bond distances, implemented on a various cloud quantum processors: 27-qubit IBM Auckland (100,000 shots per circuit), 5-qubit IBM Lima (20,000 shots per circuit), and 11-qubit IonQ (1,000 shots per circuit).}
\label{fig:12_h2mol}
\end{figure}

\section{Noise Information of Cloud Quantum Processors} \label{apx:noiseinfo}
\begin{table}[H]
\centering
\begin{tabular}{|l|l|l|l|}
\hline
Cloud Quantum Processors & IBM Auckland & IBM Lima & IonQ  \\ \hline
One Qubit Gate Times (s) & 3.55E-8 & 3.55E-8 & 1.00E-5  \\ \hline
One Qubit Gate Fidelity & 1(3)E-3 & 3(2)E-4 & 4.50E-3  \\ \hline
Two Qubit Gate Times  (s) & 5(3)E-7 & 4.1(9)E-7 & 2.00E-4  \\ \hline
Two Qubit Gate Fidelity & 1(1)E-2 & 9.51(3)E-3 & 3.86E-2  \\ \hline
Qubit Reset Time  (s) & 9.4(5)E-7 & 5.74E-6 & 2.00E-5  \\ \hline
T1 Thermal Relaxation (s)& 1.7(5)E-4 & 1.0(4)E-5 & 1.00E4  \\ \hline
T2 Dephasing (s)& 1.4(9)E-4 & 1.1(7)E-5 & 2.00E-1  \\ \hline
Readout Error & 1(1)E-2 & 3(2)E-2 & 1.30E-4  \\ \hline
State-Preparation and Measurement (SPAM) Error & NIL & NIL & 1.76E-3 \\ \hline
\end{tabular}
\caption{Qubit-Averaged Noise Parameters. Values without brackets have no published variances and are thus given to 3 significant figures.}
\label{tab:noiseinfo}
\end{table}

\section{Quantum State Fidelity}\label{apx:statefidel}

\begin{table}[H]
\centering
\begin{tabular}{|c|c|c|c|c|}
\hline
\multicolumn{5}{|c|}{Simulated IBM Auckland (100k shots/circuit)}\\ \hline 
Molecules & H$_2$  & H$_3^+$ &  LiH & H$_4$ \\ \hline
A & 0.88 & 0.56 & 0.46 & 0.24  \\ \hline
A (PS) & 0.99 & 0.88 & 0.89 & 0.70  \\ \hline
B & 0.70 & 0.50 & 0.38 & 0.23  \\ \hline
B (PS) & 0.86 & 0.81 & 0.79 & 0.63 \\ \hline
\end{tabular}

\vspace*{0.5 cm}
\centering
\begin{tabular}{|c|c|c|c|}
\hline
\multicolumn{4}{|c|}{H$_2$ Molecule}\\ \hline 
Cloud Quantum Processors & IBM Auckland  & IBM Lima &  IonQ   \\ \hline
No. of shots/circuit & 100k  & 20k  &  1k    \\ \hline
A & 0.87 & 0.80 & 0.86  \\ \hline
A (PS) & 0.98 & 0.97 & 0.97  \\ \hline
B & 0.18 & 0.24 & 0.46  \\ \hline
B (PS) & 0.48 & 0.59 & 0.77 \\ \hline
\end{tabular}
\caption{Average state fidelities of the quantum circuit sets A and B for Fig.~\ref{fig2} and Fig.~\ref{fig3}, respectively.}
\label{tab:statefidel}
\end{table}

\clearpage
\section{Post-Selection Performance}\label{apx:psperf}

\begin{table}[H]
\centering
\begin{tabular}{|c|c|c|c|c|}
\hline
\multicolumn{5}{|c|}{Simulated IBM Auckland (100k shots/circuit)}\\ \hline 
Molecules & H$_2$  & H$_3^+$ &  LiH & H$_4$ \\ \hline
A & 0.90(7) & 0.63(1) & 0.521(5) & 0.34(1)  \\ \hline
B & 0.81(4) & 0.61(2) & 0.48(1) & 0.37(2) \\ \hline
\end{tabular}

\vspace*{0.5 cm}
\centering
\begin{tabular}{|c|c|c|c|}
\hline
\multicolumn{4}{|c|}{H$_2$ Molecule}\\ \hline 
Cloud Quantum Processors & IBM Auckland  & IBM Lima &  IonQ   \\ \hline
No. of shots/circuit & 100k  & 20k  &  1k    \\ \hline
A & 0.88(8) & 0.8(1) & 0.9(1)  \\ \hline
B & 0.37(2) & 0.41(2) & 0.59(7) \\ \hline
\end{tabular}
\caption{Average post-selected fraction of measured bitstrings on the quantum circuit sets A and B for Fig.~\ref{fig2} and Fig.~\ref{fig3}, respectively.}
\label{tab:psperf}
\end{table}
\end{document}